\documentclass[twocolumn,english]{revtex4-2}
\usepackage[T1]{fontenc}
\usepackage[utf8]{inputenc}
\usepackage{color}
\usepackage{babel}
\usepackage{bm}
\usepackage{amsmath}
\usepackage{amssymb}
\usepackage{graphicx}
\usepackage[unicode=true,pdfusetitle,
 bookmarks=true,bookmarksnumbered=false,bookmarksopen=false,
 breaklinks=false,pdfborder={0 0 1},backref=false,colorlinks=true]
 {hyperref}

\makeatletter

\newcommand{\lyxdot}{.}

\renewcommand\[{\begin{equation}}
\renewcommand\]{\end{equation}}
\usepackage{xcolor}
\hypersetup{
     colorlinks   = true,
     citecolor    = blue,
     urlcolor     = blue,
     urlbordercolor = blue
}
\newcommand{\eqdef}{\equiv}

\makeatother

\usepackage[a4paper]{geometry} 
\begin{document}
\title{Arrival times of an atomic Bose-Einstein condensate}
\author{Pascal Naidon}
\email{pascal@riken.jp}

\author{Lucas Happ}
\altaffiliation{Present address: Institut für Quantenphysik and Center for Integrated Quantum Science and Technology (IQST),Universität Ulm, D-89069 Ulm, Germany}

\affiliation{Few-Body Systems Physics Laboratory, RIKEN Nishina Centre, RIKEN,
Wakō, 351-0198 Japan}
\author{Denis Boiron}
\affiliation{Université Paris-Saclay, Institut d’Optique Graduate School, CNRS,
Laboratoire Charles Fabry, 91127, Palaiseau, France}
\begin{abstract}
The times of flight of an atomic Bose-Einstein condensate are theoretically
investigated in the experimentally unexplored regime corresponding
to detection close to the trap of the condensate. In this regime,
there is no consensus on how to calculate the distribution of times
of arrival onto the detector. For non-interacting particles, distinct
theoretical predictions have been made in the past. This work analyses
how these predictions are modified for an interacting Bose-Einstein
condensate. For this purpose, a time-dependent Gross-Pitaevskii equation
is solved analytically and numerically. 
\end{abstract}
\maketitle

\section{Introduction}

Free expansions of atomic condensates are routinely performed in cold-atom
experiments as a standard technique called \emph{time-of-flight measurements}~\citep{Anderson1995,Davis1995}
to image either their density or momentum distribution. The principle
is to release the atoms forming a condensate from their trap at some
time and let them fly away until they reach some region where they
can be detected. Although in most cases only the spatial density of
the cloud is imaged, there are also experiments where the time of
flight itself is measured~\citep{Robert2001}. These measurements
are usually performed relatively far from the initial cloud of atoms.
In this regime, the times of flight can be described by a semi-classical
analysis.

It has long been pointed out that the general problem of determining
the arrival-time distribution of a particle is problematic from a
fundamental point of view. Unlike position measurement, which is well
formalised in quantum theory, time measurements are not so clearly
defined. This problem started with the realisation that there is no
self-adjoint operator for time~\citep{Pauli1958,Paul1962} and has
since been investigated by theorists for over six decades now~\citep{Muga2000}.
Several attempts have been made to predict the distribution of the
times of arrival of a non-relativistic particle on a detector. Although
all predictions agree with the semi-classical prediction in the regime
of measurements far from the initial position of the particle, they
disagree for measurements at closer distances, which we will refer
to as the\emph{ short-distance regime}. It is clearly of great interest
to test this regime experimentally in order to clarify this fundamental
point of the quantum theory.

Bose-Einstein condensates constitute a promising experimental platform
for such tests. Indeed, time-of-flight measurements are well established
in these systems, and their condensate nature allows for the measurements
of many atoms in essentially the same quantum state. However, unlike
isolated atoms prepared in the same quantum state, atoms in a Bose-Einstein
condensate interact with each other. These interactions result in
a significantly different wave function and dynamics. The goal of
the present work is to understand how the arrival times of single
atoms are modified by interactions with other atoms, and to identify
the best conditions for testing arrival-time predictions of a Bose-Einstein
condensate in the short-distance regime. The obtained results form
the basis for an experimental proposal detailed in a companion article~\citep{Naidon2025a}.

The paper is organised as follows. Section~\ref{sec:arrivals-of-isolated-particle}
presents different approaches proposed so far for predicting the arrival-time
distribution of an isolated particle. In Section~\ref{sec:arrivals-of-condensate},
these predictions are extended to a Bose-Einstein condensate within
the Gross-Pitaevskii description. Both sections are illustrated with
applications to the case of metastable helium-4 atoms.

\section{Arrivals of an isolated particle\label{sec:arrivals-of-isolated-particle}}

We consider a non-relativistic quantum particle of mass $m$ and position
$\bm{r}$, initially prepared in a state $\vert\psi_{0}\rangle$,
and subsequently evolving freely. The particle is thus described by
the wave function $\psi(\bm{r},t)$ statisfying the time-dependent
Schrödinger equation,
\begin{equation}
i\hbar\dot{\psi}(\bm{r},t)=-\frac{\hbar^{2}}{2m}\nabla^{2}\psi(\bm{r},t)\label{eq:SchroedingerEq}
\end{equation}
with the initial condition $\psi(\bm{r},0)=\psi_{0}(\bm{r})$. A detecting
surface $S$ is placed at a fixed location and registers the arrival
time of the particle. Repeating this experiment many times in the
same conditions yields different arrival times, and the question is
to detemine how these times are distributed.

\subsection{Predictions}

We consider the following predictions of the arrival-time distribution,
obtained from different theoretical approaches.

\subsubsection{The quantum clock approach}

Since the Born rule interprets $\left|\psi(\bm{r},t)\right|^{2}$
at a fixed time~$t$ as the probability density for finding the particle
at position $\bm{r}$, perhaps the most natural expectation for a
general practioner of quantum physics is that the same quantity at
a fixed position $\bm{r}$ gives the distribution of times $t$ at
which the particle is to be detected at that location. It is indeed
the conclusion reached by the quantum clock approach~\citep{Giovannetti2015,Maccone2020,Roncallo2023}
which is based on the idea that time itself should be treated quantum-mechanically,
as the observed position of a clock hand that is entangled with the
particle~\citep{Page1983,Aharonov1984}. This approach predicts that
the distribution $\Pi(t)$ of times $t$ for the particle to arrive
onto a surface $S$ is simply proportional to its probability of presence
on the surface 
\begin{equation}
\Pi(t)\eqdef\alpha\iint_{S}\vert\psi(\bm{r},t)\vert^{2}dS.\label{eq:densityFlux}
\end{equation}
Note however that the factor $\alpha$, homogeneous to a velocity,
is not specified by the theory, and might depend on the detector's
characteristics. We call the predicted distribution Eq.~(\ref{eq:densityFlux})
the \emph{density-like flux}.

\subsubsection{The axiomatic approach}

Most of the early works on the arrival-time problem arrived at a predicted
distribution~\citep{Aharonov1961,Allcock1969b,Kijowski1974,Grot1996,Delgado1997,Anastopoulos2006},
often called the \emph{Kijowski distribution}~\citep{Kijowski1974,Muga2000}.
This distribution was obtained either by an axiomatic approach~\citep{Kijowski1974}
(using general principles of quantum and classical mechanics) or by
constructing a reasonable operator for the time of arrival~\citep{Aharonov1961}.
For these reasons, it has been claimed to represent the prediction
of standard quantum theory~\citep{Aharonov1961,Allcock1969a,Allcock1969b,Allcock1969c,Kijowski1974,Werner1986,Grot1996,Delgado1997,Aharonov1998,Baute2000,Galapon2005,Juric2022},
although this has been disputed~\citep{Kijowski1999,Leavens2002,Egusquiza2003,Leavens2005}.

Formally, the distribution consists of a sum of two terms, one for
momenta pointing towards one side of the surface $S$ and the other
for momenta pointing towards the opposite side. For a planar surface
perpendicular to the $x$-axis and located at some coordinate $x_{S}$,
it reads~\citep{Kijowski1974,Muga2000},
\begin{multline}
\Pi_{\text{K}}(t)\!=\!\frac{\hbar}{m}\sum_{\pm}\!\!\iint\!\frac{dk_{y}}{2\pi}\frac{dk_{z}}{2\pi}\times\\
\left|\int_{-\infty}^{\infty}\!\!\!\!\!dk_{x}\Theta(\pm k_{x})\left|k_{x}\right|^{\!\frac{1}{2}}e^{ik_{x}x_{S}}\tilde{\psi}(\bm{k},t)\right|^{2},\label{eq:KijowskiDistribution-MomentumRepresentation}
\end{multline}
where $\tilde{\psi}(\bm{k},t)\eqdef\int d\bm{r}e^{-i\bm{k}\cdot\bm{r}}\psi(\bm{r},t)$
is the Fourier transform of $\psi$, and $\Theta$ is the Heaviside
step function. It can be expressed in terms of $\psi$ as follows\,\citep{Leavens2005},
\begin{multline}
\Pi_{\text{K}}(t)\eqdef\frac{\hbar}{32\pi m}\sum_{\pm}\iint_{S}dydz\times\\
\left|\int_{-\infty}^{\infty}dx\left(\psi(\bm{r},t)-\psi(x_{S},y,z,t)\right)\frac{1\pm i\,\text{sign}(x-x_{S})}{\left|x-x_{S}\right|^{3/2}}\right|^{2}.\label{eq:KijowskiDistribution}
\end{multline}
Similar formulae are obtained for other orientations of $S$ by
permuting $x$,$y$, and $z$. We will refer to this prediction as
the \emph{Kijowski flux}.

\subsubsection{The quantum flux approach}

Another prediction for the arrival-time distribution derives from
the hydrodynamic interpretation~\citep{Madelung1927} of the Schrödinger
equation, where the probability density $\rho(\bm{r},t)=\vert\psi(\bm{r},t)\vert^{2}$
flows following the probability current, $\bm{j}=\frac{\hbar}{m}\text{Im}\left(\psi^{*}\bm{\nabla}\psi\right)$,
through the continuity equation $\dot{\rho}+\bm{\nabla}\cdot\bm{j}=0$.
According to this description, the distribution of times $t$ for
the particle to arrive onto some surface $S$ is the flux 
\begin{equation}
F(t)\eqdef\iint_{S}\bm{j}(\bm{r},t)\cdot d\bm{S}\label{eq:probabilityFlux}
\end{equation}
of the probability current through the detecting surface, whose normal
vector is oriented towards the inside of the detector. As long as
$F$ remains positive, this description is consistent~\citep{McKinnon1995,Leavens1998}
with the Bohmian interpretation~\citep{Bohm1952,Bohm1952a} of quantum
mechanics (also known as de Broglie-Bohm's pilot wave theory), according
to which the particle has a definite trajectory that follows the probability
current. We will call the predicted distribution Eq.~(\ref{eq:probabilityFlux})
the \emph{probability flux}.

\subsubsection{The stochastic path approach}

Nelson's stochastic mechanics~\citep{Fenyes1952,Nelson1966,Nelson1985}
also proposes that the particle follows a definite trajectory. However,
instead of being smooth and deterministic like in the pilot wave theory,
the trajectory is stochastic, i.e. non-differentiable and undeterministic,
like Brownian motion. It was shown~\citep{Naidon2024} that the arrival-time
distribution of such trajectories is approximately given by the flux
\begin{equation}
F_{-}(t)\eqdef\iint_{S}\bm{j}_{-}(\bm{r},t)\cdot d\bm{S}\label{eq:backwardFlux}
\end{equation}
 of the backward current $\bm{j}_{-}=\bm{j}-\bm{i}$, where $\bm{i}=\frac{\hbar}{m}\text{Re}\left(\psi^{*}\bm{\nabla}\psi\right)$.
We call this predicted distribution the \emph{backward flux}. Note
that this result was obtained assuming that the particle is detected
on its first passage into the detector. In the opposite limit where
the particle can enter the detector many times before being detected,
the predicted arrival-time distribution was shown~\citep{Nitta2008,Naidon2024}
to be identical to the density-like flux Eq.~(\ref{eq:densityFlux}).

\subsubsection{The semi-classical approximation}

The previous distributions are often compared with the so-called semi-classical
approximation, which has been in good agreement with all arrival-time
measurements done so far. The idea behind this approximation is that
the arrival-time measurement is regarded as a quantum measurement
of the initial momentum $\hbar k_{x}$ of the particle in the direction
$x$ towards the detecting surface, and the particle is assumed to
travel classically with that initial momentum until it reaches the
surface. The initial distribution of momenta $\hbar k_{x}$ along
the $x$ direction is given by
\[
n(k_{x})=\iint\frac{dk_{y}}{2\pi}\frac{dk_{z}}{2\pi}\left|\tilde{\psi}_{0}(k_{x},k_{y},k_{z})\right|^{2}
\]
where 
\[
\tilde{\psi}_{0}(\bm{k})\eqdef\int d^{3}\bm{r}\psi_{0}(\bm{r})e^{-i\bm{k}\cdot\bm{r}}
\]
is the Fourier transform of the initial wave function $\psi_{0}(\bm{r})$.
Assuming that the surface is planar and separated from the centre
of the initial wave function by a distance $\Delta x$ that is much
larger than the spatial extent of that initial wave function, then
the time it takes for a classical particle to reach it is $m\Delta x/(\hbar k_{x})$.
One thus arrives at the semi-classical arrival-time distribution,
\begin{equation}
\Pi_{\text{SC}}(t)\eqdef\frac{m\Delta x}{2\pi\hbar t^{2}}\iint\frac{dk_{y}dk_{z}}{(2\pi)^{2}}\left|\tilde{\psi}_{0}\left(\frac{m}{\hbar}\frac{\Delta x}{t},k_{y},k_{z}\right)\right|^{2},\label{eq:SC-flux}
\end{equation}
which by construction can only be valid when the detection occurs
far from the initial location of the particle.

\subsection{Effect of gravity}

\subsubsection{Frame transformation}

The above predictions Eqs.~(\ref{eq:densityFlux}-\ref{eq:backwardFlux})
are applicable to a free particle. However, in a terrestrial laboratory,
even in vacuum and in the absence of any externally applied force,
the particle is subject to the Earth's gravity along the $z$-axis.
It is straightforward to obtain the predictions in the presence of
gravity: one can relate the freely-evolving wave function $\psi$
in a falling frame to the wave function $\psi_{\text{lab}}$ in the
laboratory frame through the transformation~\citep{Kennard1927,Zimmermann2017}
\begin{equation}
\psi_{\text{lab}}(\bm{r},t)\eqdef\psi(\bm{r}-\bm{r}_{\text{fall}}(t),t)e^{i\varphi(\bm{r},t)}\label{eq:Lab-frame-wavefunction}
\end{equation}
where the phase factor $\varphi$ is given by
\begin{equation}
\varphi(\bm{r},t)\eqdef-\frac{mt}{\hbar}\left(gz+\frac{g^{2}t^{2}}{6}\right)\label{eq:gravitational-phase}
\end{equation}
with $g$ being the Earth's gravitational acceleration, and $\bm{r}_{\text{fall}}(t)$
being the classical falling motion,
\begin{equation}
\bm{r}_{\text{fall}}(t)\eqdef-\frac{1}{2}gt^{2}\bm{e}_{z}\label{eq:Classical-Falling-Trajectory}
\end{equation}
where $\bm{e}_{z}$ is the up-pointing vertical unit vector.

Since gravity prominently affects arrivals along the vertical direction,
in the following we will consider a horizontal detecting surface $S$,
set at some vertical coordinate $z_{S}\eqdef-\Delta z$.

\subsubsection{Falling density and fluxes}

By replacing $\psi$ by $\psi_{\text{lab}}$ in Eqs.~(\ref{eq:densityFlux}-\ref{eq:backwardFlux}),
one obtains the predicted arrival-time distributions for a free falling
particle. Although the phase factor $\varphi$ does not change observables
such as densities, it does affect fluxes such as Eq.~(\ref{eq:probabilityFlux})
and (\ref{eq:backwardFlux}). Specifically, one finds:
\begin{align}
\Pi_{\text{lab}}(t) & =\alpha\iint_{S}\vert\psi(\bm{r}-\bm{r}_{\text{fall}}(t),t)\vert^{2}dS\label{eq:fallDensityFlux}\\
F_{\text{lab}}(t) & =\iint_{S}\bm{j}(\bm{r}-\bm{r}_{\text{fall}}(t),t)\cdot d\bm{S}+F_{\text{fall}}(t)\label{eq:fallProbabilityFlux}\\
F_{-,\text{lab}}(t) & =\iint_{S}\bm{j}_{-}(\bm{r}-\bm{r}_{\text{fall}}(t),t)\cdot d\bm{S}+F_{\text{fall}}(t)\label{eq:FallBackwardFlux}
\end{align}
with the flux
\begin{equation}
F_{\text{fall}}(t)=-\iint_{S}\vert\psi(\bm{r}-\bm{r}_{\text{fall}}(t),t)\vert^{2}gt\bm{e}_{z}\cdot d\bm{S}\label{eq:fall-F}
\end{equation}
which is proportional to $t$, and thus becomes the dominant contribution
for large times.

\subsubsection{Falling Kijowski flux}

The Kijowski flux $\Pi_{\text{K,lab}}$ in the presence of gravity
is not directly connected to the distribution $\Pi_{\text{K}}$ in
the absence of gravity. However, one can find its limit for large
times.  Using the frame transformation Eqs.~(\ref{eq:Lab-frame-wavefunction}-\ref{eq:Classical-Falling-Trajectory}),
one finds
\begin{equation}
\tilde{\psi}_{\text{lab}}(\bm{k},t)=\tilde{\psi}(\bm{k}+\frac{m}{\hbar}gt\bm{e}_{z},t)e^{i\frac{1}{2}gt^{2}\left(k_{z}+\frac{2}{3}\frac{m}{\hbar}gt\right)},\label{eq:fall-psi-MomentumRepresentation}
\end{equation}
which can be inserted into the formula of Eq.~(\ref{eq:KijowskiDistribution-MomentumRepresentation})
for a horizontal detecting surface displaced along the $z$-axis.
This gives, after a change of integration variable:
\begin{align}
 & \Pi_{\text{K,lab}}(t)=\frac{\hbar}{m}\!\int\!\!\!\!\int_{S}\!\frac{dk_{x}}{2\pi}\frac{dk_{y}}{2\pi}\times\nonumber \\
 & \Bigg(\left|\int_{-\infty}^{\frac{m}{\hbar}gt}\!\!\!\!\!dk_{z}\left|k_{z}\!-\!\frac{m}{\hbar}gt\right|^{\!\frac{1}{2}}\tilde{\psi}(\bm{k},t)e^{ik_{z}\left(z_{S}+\frac{1}{2}gt^{2}\right)}\right|^{2}+\nonumber \\
 & \quad\left|\int_{\frac{m}{\hbar}gt}^{\infty}\!\!\!\!\!dk_{z}\left|k_{z}\!-\!\frac{m}{\hbar}gt\right|^{\!\frac{1}{2}}\tilde{\psi}(\bm{k},t)e^{ik_{z}\left(z_{S}+\frac{1}{2}gt^{2}\right)}\right|^{2}\Bigg).\label{eq:Pi_K-lab-intermediate-calculation}
\end{align}

Since $\tilde{\psi}$ has a support $\left|k_{z}\right|<k_{\text{max}}$
outside which it is essentially zero, for sufficiently long times
$t$ such that $mgt/\hbar\gg k_{\text{max}}$, the second integral
over $k_{z}$ vanishes, and the term $mgt/\hbar$ dominates over $k_{z}$
in the first integral, so that it can be moved out. This results in
\begin{equation}
\Pi_{\text{K,lab}}(t)\xrightarrow[t\to\infty]{}gt\iint_{S}dxdy\left|\psi(x,y,-\Delta z+\frac{1}{2}gt^{2},t)\right|^{2}\label{eq:fall-Kijowski}
\end{equation}
which is the same quantity as the flux of Eqs.~(\ref{eq:fall-F}).

\subsubsection{Falling semi-classical approximation}

To use the semi-classical approximation in the presence of gravity,
one again starts from the initial momentum distribution, and assumes
the particle follows the classical trajectory $z(t)=z_{0}+\frac{\hbar}{m}k_{z}t-\frac{1}{2}gt^{2}$
along the $z$ direction, taking into account both the initial momentum
$\hbar k_{z}$ and the gravitional pull $g$. This gives the following
distribution,
\begin{multline}
\Pi_{\text{SC}}(t)=\frac{m}{2\pi\hbar}\left(\!\frac{\Delta z}{t^{2}}\!+\!\frac{1}{2}g\!\right)\times\\
\iint\frac{dk_{x}}{2\pi}\frac{dk_{y}}{2\pi}\left|\tilde{\psi}\left(\!k_{x},k_{y},\frac{m}{\hbar}\left(\!-\frac{\Delta z}{t}\!+\!\frac{1}{2}gt\!\right)\!\right)\right|^{2}.\label{eq:fall-SC-flux}
\end{multline}

\subsubsection{Gravity-dominated regime\label{subsec:Fall-dominated-regime}}

For large times, all predictions {[}Eqs.~(\ref{eq:fallDensityFlux}-\ref{eq:FallBackwardFlux})
and (\ref{eq:fall-Kijowski}){]} become proportional to the ``falling
flux'' Eq.~(\ref{eq:fall-F}), which can be interpreted classically
as the flux of a density $\left|\psi\right|^{2}$ with falling speed
$gt$. We call this situation the ``gravity-dominated regime''.
Although the situation is less obvious for the semi-classical distribution
Eq.~(\ref{eq:fall-SC-flux}), we will see in Sec.~\ref{subsec:Gaussian-wave-packet}
that it also tends to the falling flux in the case of a Gaussian wave
packet.

One can estimate the typical time beyond which all predictions converge
to the gravity-dominated regime. For a wave packet $\psi$ of typical
size $\sigma$, the typical speed of expansion is $\hbar/m\sigma$.
This speed is dominated by the falling speed $gt$ for times much
larger than 
\begin{equation}
t_{\sigma}\eqdef\frac{\hbar}{mg\sigma}.\label{eq:t_sigma}
\end{equation}
The same time scale is obtained in the derivation of Eq.~(\ref{eq:fall-Kijowski})
through the condition $mgt/\hbar\gg k_{\text{max}}$ with $k_{\text{max}}\sim1/\sigma$.

In order to distinguish the different predicted arrival-time distributions
of a free-falling particle, the typical time to reach the detector
given by the classical fall time
\begin{equation}
t_{\text{fall}}\eqdef\sqrt{2\Delta z/g}\label{eq:t_fall}
\end{equation}
should not be greater than $t_{\sigma}$. Moreover, in order not
to perturb the initial state of the particle, the detector should
be placed sufficiently far from --- or at the very least outside
--- the region where the particle is initially localised, i.e. $\Delta z>\sigma$.
This leads to the condition $\sigma<\sigma_{g}$, with
\begin{equation}
\sigma_{g}\eqdef\left(\frac{\hbar^{2}}{2m^{2}g}\right)^{1/3}.\label{eq:sigma_g}
\end{equation}
For a helium-4 atom, this requires a trapping size smaller than $\sigma_{g}\approx2\,\mu\text{m}$.

\subsection{Case of a Gaussian wave packet\label{subsec:Gaussian-wave-packet}}

All the above predicted distributions can be calculated analytically
in the case of a freely expanding wave packet $\psi$ that is initially
Gaussian,
\[
\psi_{0}(\bm{x})=\frac{1}{\pi^{3/2}\left(\sigma_{x}\sigma_{y}\sigma_{z}\right)^{1/2}}\exp\left(-\frac{x^{2}}{2\sigma_{x}^{2}}-\frac{y^{2}}{2\sigma_{y}^{2}}-\frac{z^{2}}{2\sigma_{z}^{2}}\right).
\]
This situation corresponds physically to an isolated particle prepared
in the ground state of a harmonic trap 
\begin{equation}
V(\bm{x})=\frac{1}{2}m\sum_{i}\omega_{i}^{2}x_{i}^{2}\label{eq:HarmonicTrap}
\end{equation}
 with frequencies $\omega_{i}\eqdef\hbar/m\sigma_{i}^{2}$, which
is suddenly released at time $t=0$. Note that in this case the motions
along $x$, $y$, and $z$ are independent.

\begin{figure}
\includegraphics[height=7cm]{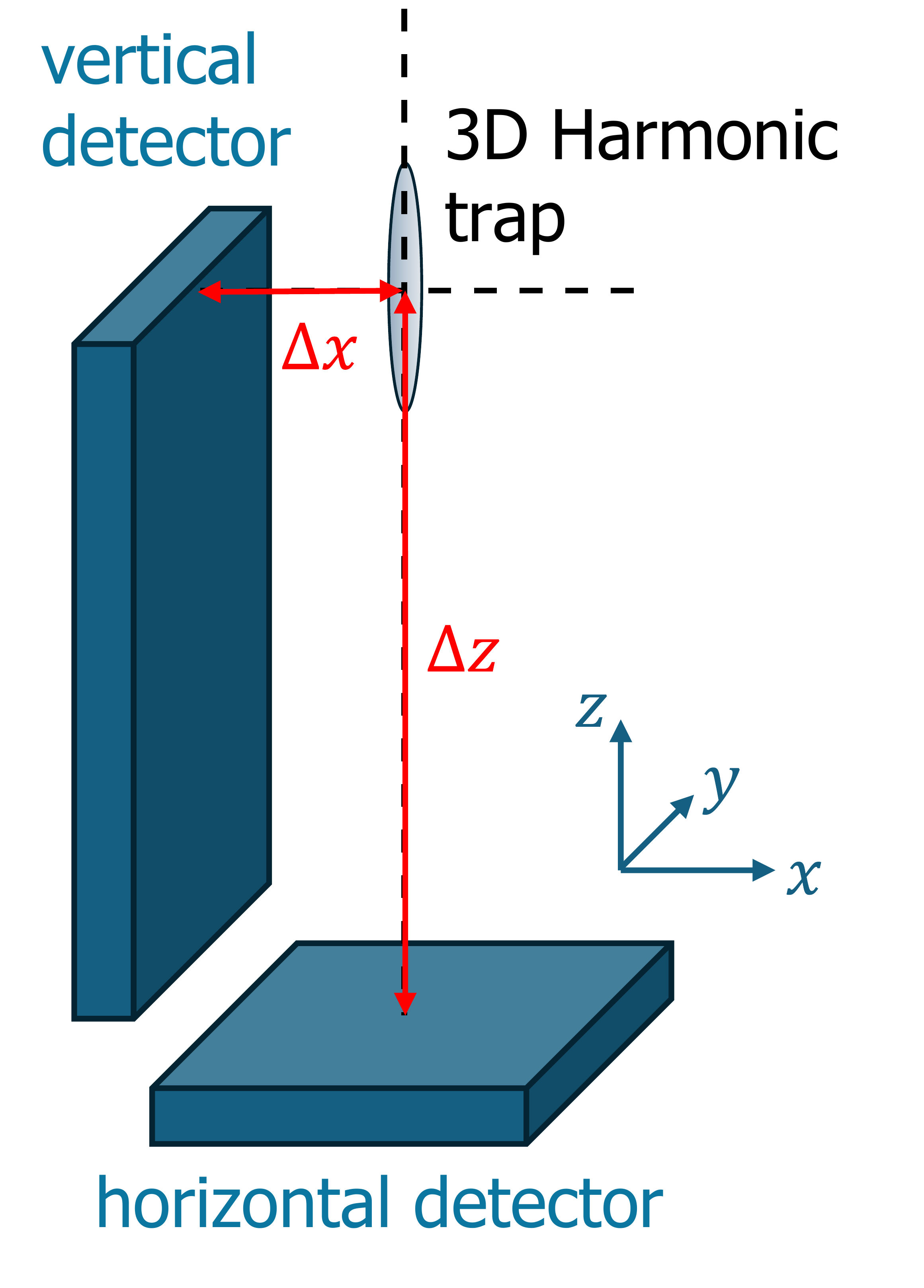}

\caption{\label{fig:setup-single-atom}Schematic setup for measuring the arrival
times of an atomic Gaussian wave packet. A metastable atom is initially
prepared in the ground state of a three-dimensional harmonic trap
(shown here as an ellipsoid). The atom is then released from the trap
and observed on a horizontal detector and a vertical detector separated
from the trap by distances $\Delta z$ and $\Delta x$ respectively,
as indicated by the red arrows.}
\end{figure}

The arrivals along $z$ and $x$ can be independently monitored by
placing horizontal and vertical planar detectors at some distances
$\Delta z$ and $\Delta x$ from the centre of the trap, as depicted
in Fig.~\ref{fig:setup-single-atom}. The spatial extent of the wave
function $\psi$ is assumed to remain small compared to the size of
the detectors, which is justified in all considered scenarios, so
that the detectors can be regarded as infinite planar surfaces.

\subsubsection{Analytical formulae}

For the horizontal detector, one finds the arrival distributions:
\begin{align}
\Pi(t) & =\alpha\frac{1}{\sqrt{\pi}\sigma_{z}\lambda_{z}(t)}\exp\left(-\frac{\Delta z^{\prime2}}{\sigma_{z}^{2}\lambda_{z}(t)^{2}}\right)\label{eq:Gaussian-DensityFlux-Falling}\\
\Pi_{\text{K}}(t) & =\frac{\omega_{z}}{4}\frac{\sqrt{\pi}e^{-\left(\frac{gt}{\sigma_{z}\omega_{z}}\right)^{2}}}{\lambda_{z}(t)^{3/2}}\sum_{\pm}\left|K_{\pm}\left(s\right)\right|^{2}\label{eq:Gaussian-KijowskiFlux-Falling}\\
F(t) & =\left(\frac{\omega_{z}^{2}\Delta z^{\prime}}{\lambda_{z}(t)^{2}}-g\right)t\frac{e^{-\frac{\Delta z^{\prime2}}{\sigma_{z}^{2}\lambda_{z}(t)^{2}}}}{\sqrt{\pi}\sigma_{z}\lambda_{z}(t)}\label{eq:Gaussian-ProbabilityFlux-Falling}\\
F_{-}(t) & =\left(\frac{\omega_{z}\left(\omega_{z}t+1\right)\Delta z^{\prime}}{\lambda_{z}(t)^{2}}-gt\right)\frac{e^{-\frac{\Delta z^{\prime2}}{\sigma_{z}^{2}\lambda_{z}(t)^{2}}}}{\sqrt{\pi}\sigma_{z}\lambda_{z}(t)}\label{eq:Gaussian-BackwardFlux-Falling}\\
\Pi_{\text{SC}}(t) & =\frac{1}{\omega_{z}}\left(\frac{\Delta z^{\prime}}{t^{2}}+g\right)\frac{\exp\left(-\frac{\Delta z^{\prime2}}{\sigma_{z}^{2}\omega_{z}^{2}t^{2}}\right)}{\sqrt{\pi}\sigma_{x}}\label{eq:Gaussian-Semi-classical-Falling}
\end{align}
where
\begin{equation}
\lambda_{z}(t)\eqdef\sqrt{1+\omega_{z}^{2}t^{2}}\xrightarrow[t\to\infty]{}\omega_{z}t\label{eq:Scaling-Factor}
\end{equation}
is the time-dependent scaling factor accounting for the expansion
of the wave packet in the $z$ direction, and
\begin{align}
\Delta z^{\prime} & \eqdef\Delta z-\frac{1}{2}gt^{2}\label{eq:Deltaz_prime}\\
s & \eqdef-\frac{\left(\left(\frac{gt}{\omega_{z}}(i-\omega_{z}t)-\Delta z^{\prime}\right)/2\sigma_{z}\right)^{2}}{1+i\omega_{z}t}.\label{eq:s}
\end{align}

For the vertical detector, one finds the arrival-time distributions:
\begin{align}
\Pi(t) & =\alpha\frac{1}{\sqrt{\pi}\sigma_{x}\lambda_{x}(t)}\exp\left(-\frac{\Delta x^{2}}{\sigma_{x}^{2}\lambda_{x}(t)^{2}}\right)\label{eq:Gaussian-DensityFlux}\\
\Pi_{\text{K}}(t) & =\frac{\omega_{x}}{4}\frac{\sqrt{\pi}}{\lambda_{x}(t)^{3/2}}\sum_{\pm}\left|K_{\pm}\left(-\frac{\left(\Delta x/2\sigma_{x}\right)^{2}}{1+i\omega_{x}t}\right)\right|^{2}\label{eq:Gaussian-KijowskiFlux}\\
F(t) & =\frac{\omega_{x}^{2}t}{\lambda_{x}(t)^{3}}\frac{\Delta x}{\sqrt{\pi}\sigma_{x}}\exp\left(-\frac{\Delta x^{2}}{\sigma_{x}^{2}\lambda_{x}(t)^{2}}\right)\label{eq:Gaussian-ProbabilityFlux}\\
F_{-}(t) & =\frac{\omega_{x}\left(\omega_{x}t+1\right)}{\lambda_{x}(t)^{3}}\frac{\Delta x}{\sqrt{\pi}\sigma_{x}}\exp\left(-\frac{\Delta x^{2}}{\sigma_{x}^{2}\lambda_{x}(t)^{2}}\right)\label{eq:Gaussian-BackwardFlux}\\
\Pi_{\text{SC}}(t) & =\frac{1}{\omega_{x}t^{2}}\frac{\Delta x}{\sqrt{\pi}\sigma_{x}}\exp\left(-\left(\frac{\Delta x}{\sigma_{x}\omega_{x}t}\right)^{2}\right)\label{eq:Gaussian-Semi-classical}
\end{align}
Since there is no influence of gravity in this case, one can check
that Eqs.~(\ref{eq:Gaussian-DensityFlux}-\ref{eq:Gaussian-Semi-classical})
can be obtained by setting $g\to0$ and $z\to x$ in Eqs.~(\ref{eq:Gaussian-DensityFlux-Falling}-\ref{eq:Gaussian-Semi-classical-Falling}).

The function $K_{\pm}$ in Eqs.~(\ref{eq:Gaussian-KijowskiFlux-Falling})
and (\ref{eq:Gaussian-KijowskiFlux}) reads
\begin{equation}
K_{\pm}\!(s)=\frac{e^{s}}{s^{1/4}}\!\left[s\!\left(I_{\!\frac{5}{4}}(s)\pm I_{\!\frac{3}{4}}(s)\pm I_{\!-\frac{1}{4}}(s)\right)\!+\!\left(s\!+\!\tfrac{1}{2}\right)\!I_{\!\frac{1}{4}}(s)\right].\label{eq:FunctionK}
\end{equation}
where $I_{\alpha}(z)$ denotes the modified Bessel function of the
first kind. To our knowledge, this is the first time such explicit
analytic expressions for the Kijowski flux have been given.

\subsubsection{Arrivals on a horizontal detector\label{subsec:Vertical-arrivals}}

\begin{figure}
\includegraphics[width=8cm]{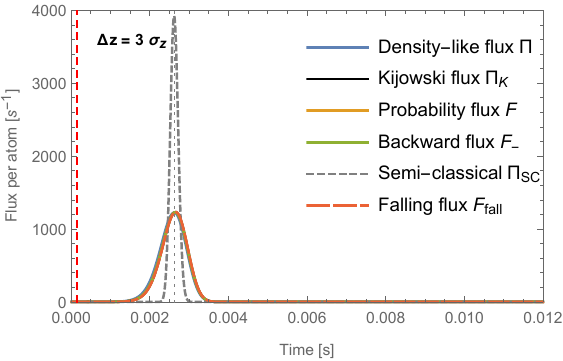}

\includegraphics[width=8cm]{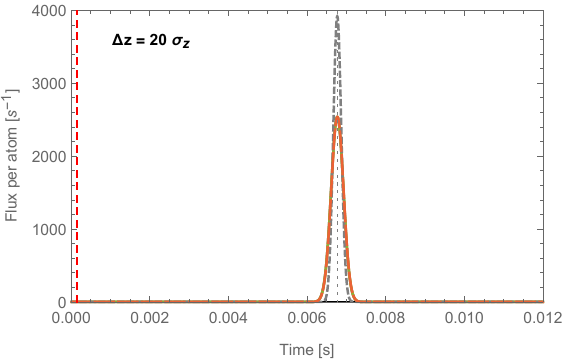}

\includegraphics[width=8cm]{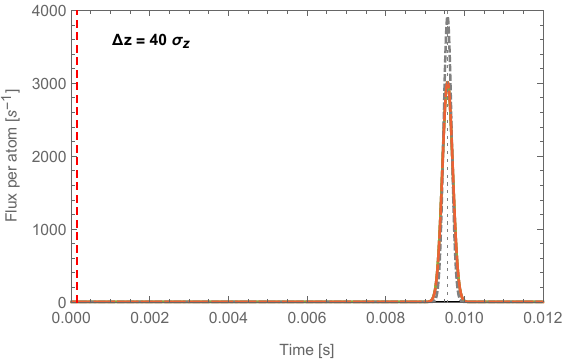}

\caption{\label{fig:fall150Hz-Gaussian}Arrival-time distributions for an isolated
helium-4 atom detected onto a horizontal planar detector, based on
different predictions Eqs.~(\ref{eq:Gaussian-DensityFlux-Falling}-\ref{eq:Gaussian-Semi-classical-Falling}).
For the density-like flux Eq.~(\ref{eq:Gaussian-DensityFlux-Falling}),
the coefficient $\alpha$ is set to $gt_{\text{fall}}$. The atom
is initially trapped in a harmonic trap with frequency $\omega_{z}=2\pi\times20\,\text{Hz}$
and the detector is separated from the trap by a distance $\Delta z=3,\,20,\,40\,\sigma_{x}$,
(from top to bottom panel) corresponding approximately to \textcolor{blue}{${\normalcolor 0.03,\,0.2,\,0.4\,\text{mm}}$}.
The gray vertical dotted line indicates the classical fall time $t_{\text{fall}}$
Eq.~(\ref{eq:t_fall}), and the red vertical dashed line the time
$t_{\sigma}$ Eq.~(\ref{eq:t_sigma}) beyond which the system is
in the gravity-dominated regime. All predictions  are nearly indistinguishable
and well reproduced by the falling flux Eq.~(\ref{eq:fall-F}) shown
by the red long-dashed curve.}
\end{figure}

Let us first consider the vertical arrivals onto the horizontal detector.
The distributions Eqs.~(\ref{eq:fallDensityFlux}-\ref{eq:FallBackwardFlux})
and (\ref{eq:fall-Kijowski}-\ref{eq:fall-SC-flux}) are plotted in
Fig.~\ref{fig:fall150Hz-Gaussian} for an isolated helium-4 atom
initially held in a trap of frequency $\omega_{z}=2\pi\times20\,\text{Hz}$
for different detector positions $\Delta z$. The trapping size $\sigma_{z}=(\hbar/m\omega_{z})^{1/2}$
is about $11\,\mu\text{m}$, which is larger than $\sigma_{g}\approx2\,\mu\text{m}$
of Eq.~(\ref{eq:sigma_g}). As expected from the discussion in Sec.~\ref{subsec:Fall-dominated-regime},
the system is in the gravity-dominated regime and all predictions
are nearly indistinguishable from the falling flux Eq.~(\ref{eq:fall-F}),
shown by the red dashed line. The semi-classical approximation (grey
dashed line) is noticeably different when the detector is close to
the trap but approaches the falling flux for larger distances.

\begin{figure}
\includegraphics[width=8cm]{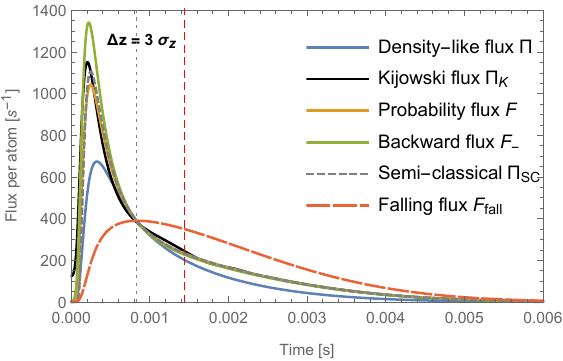}

\includegraphics[width=8cm]{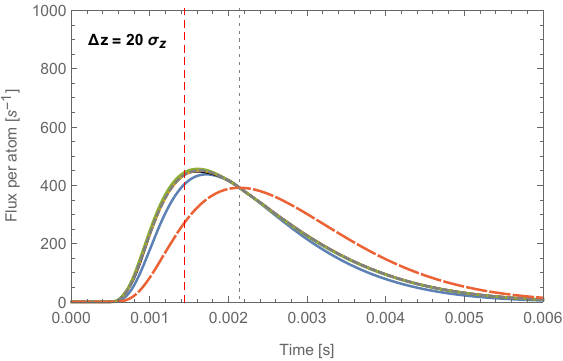}

\includegraphics[width=8cm]{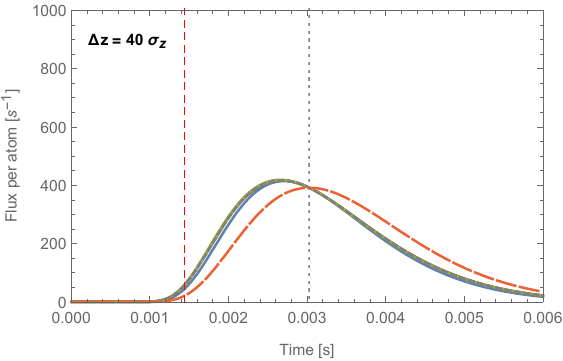}

\caption{\label{fig:fall4000Hz-Gaussian}Same as Fig.~\ref{fig:fall150Hz-Gaussian}
for a trapping frequency $\omega_{z}=2\pi\times2000\,\text{Hz}$.
The distance of the detector is still $\Delta z=3,\,20,\,40\,\sigma_{x}$,
(from top to bottom panel) which now corresponds to approximately
$0.003,\,0.022,\,0.044\,\text{mm}$. For $\Delta z\ge20\sigma_{x}$,
all predictions (solid curves) are nearly indistinguishable and well
reproduced by the semi-classical flux Eq.~(\ref{eq:Gaussian-Semi-classical-Falling})
shown by the gray dashed curve. }
\end{figure}

In an effort to avoid the gravity-dominated regime, we consider a
much tighter trap, with trapping frequency $\omega_{z}=2\pi\times2000\,\text{Hz}$,
corresponding to a trapping size $\sigma_{z}$ of about $1\,\mu\text{m}$,
smaller than the value of $\sigma_{g}$. The corresponding arrival-time
distributions are shown in Fig.~\ref{fig:fall4000Hz-Gaussian}. Now,
one can see that the predictions exhibit some noticeable differences
when the detector is very close to the trap (top panel, where $\Delta z$
is only three times the size $\sigma_{z}$, i.e. about $3\,\mu\text{m}$).
Nevertheless, these differences are mostly quantitative and would
be difficult to distinguish experimentally. For larger distances of
the detector, the different predictions quickly become indistinguishable.
In these tighter trapping conditions, it is the semi-classical approximation
(grey dashed curve) that accurately reproduces to the predictions
rather than the falling flux (red long-dashed curve). The falling
flux approaches the predictions only for large distances between the
trap and the detector, corresponding to times larger than $t_{\sigma}$
(vertical red dashed line)..

From these rather unfavourable results, and considering that producing
such tight traps and placing a detector at such close distances would
be technically challenging, we conclude that measuring vertical arrivals
under the influence of gravity shows no promise of discriminating
theories. A radical solution would be to perform the experiment in
low-gravity environments, such as the International Space Station
where Bose-Einstein condensates have been realised before~\citep{Aveline2020}.
A more mundane approach is to measure arrival times in horizontal
directions, which we will now turn to.

\subsubsection{Arrivals on a vertical detector}

\begin{figure}
\includegraphics[width=8cm]{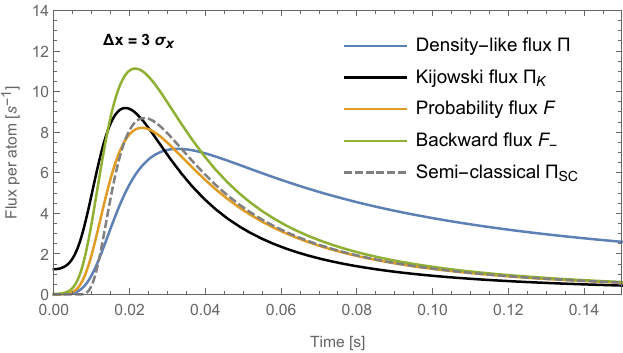}

\includegraphics[width=8cm]{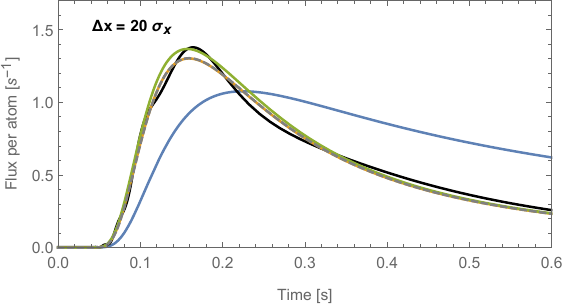}

\includegraphics[width=8cm]{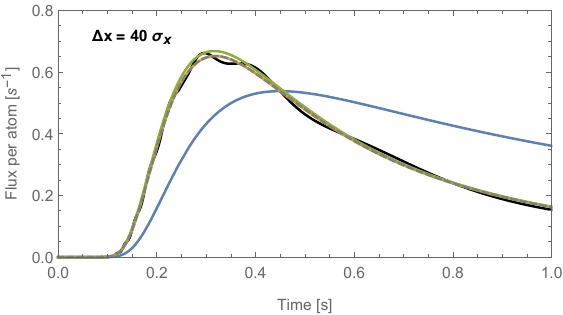}

\includegraphics[width=8cm]{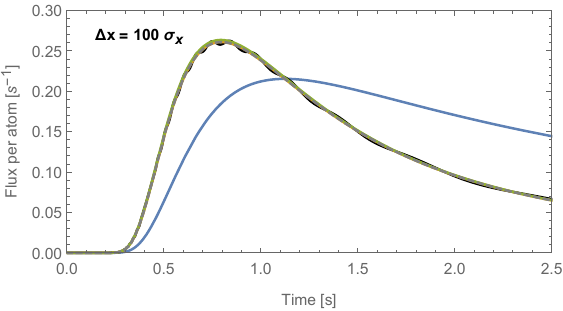}

\caption{\label{fig:Gaussian} Arrival-time distributions for an isolated helium-4
atom detected onto a vertical planar detector, based on the different
predictions, Eqs.~(\ref{eq:Gaussian-DensityFlux}-\ref{eq:Gaussian-Semi-classical}).
The atom is initially trapped in a harmonic trap with frequency $\omega_{x}=2\pi\times20\,\text{Hz}$
and the detector is separated from the trap by a distance $\Delta x=3,\,20,\,40,\,100\,\sigma_{x}$,
(from top to bottom panel) corresponding approximately to $0.03,\,0.2,\,0.4,\,1.1\,\text{mm}$.
Only the density-like flux (blue curve) differs significantly from
the other predictions for large $\Delta x$. It is given by the probability
density scaled by an arbitrary coefficient $\alpha$ set here to\textcolor{blue}{{}
}1.0 mm/s.}
\end{figure}

Let us now consider horizontal arrivals on the vertical detector,
which are not directlty affected by gravity. The distributions of
Eqs.~(\ref{eq:Gaussian-DensityFlux}-\ref{eq:Gaussian-Semi-classical})
are illustrated in Fig.~\ref{fig:Gaussian} for an isolated helium-4
atom initially held in a trap of frequency $\omega_{x}=2\pi\times20\,\text{Hz}$,
for different detector positions $\Delta x$. Note that there is now
only one scale in this problem, the initial extent $\sigma_{x}$ of
the wave packet, which is directly related to the trapping frequency
$\omega_{x}=\hbar/m\sigma_{x}^{2}$, so that results for other values
of $\omega_{x}$ would look identical upon a proper rescaling of units.
One an see from Fig.~\ref{fig:Gaussian} that these distributions
differ significantly only when $\Delta x$ is of the order of $\sigma_{x}$.
As $\Delta x$ increases, all distributions rapidly approach one another,
except the density-like flux, which remains markedly different, as
discussed before in the literature~\citep{Roncallo2023,Cavendish2024,Maccone2025}.

In addition, one can notice that the Kijowski flux $\Pi_{\text{K}}$
exhibits for large separations $\Delta x$ a faint but noticeable
oscillatory pattern, whose frequency increases with increasing separation
$\Delta x$. We also note that the amplitude of these oscillations
appears to be decreasing with increasing $\Delta x$. We are not aware
of any previous discussion of these oscillations. We checked that
they are not an artefact of numerical calculations, since they appear
both in direct numerical calculations of Eq.~(\ref{eq:KijowskiDistribution})
and in calculations using the exact analytic form Eq.~(\ref{eq:Gaussian-KijowskiFlux}).
It is not clear from a physical point of view why these oscillations
would occur, but the mathematical structure of Eq.~(\ref{eq:KijowskiDistribution}),
similar to a Fourier transform, undoubtedly produces them. Since they
are absent from other predictions Eqs.~(\ref{eq:densityFlux}-\ref{eq:backwardFlux}),
they constitute a telltale sign that could discriminate the Kijowski
flux from other predictions.

Although the arrival-time measurement of a single trapped atom could
in principle be performed, a very large number of measurements would
be necessary to obtain sufficient statistics for resolving the oscillations.
It is much more practical to perform the experiment with a condensate
of many identical atoms. Then come the questions of how the arrival-time
distributions are modified in a Bose-Einstein condensate, and whether
the oscillations in the Kijowski flux are still present, which will
be investigated in the next section.

\section{Arrivals of a condensate\label{sec:arrivals-of-condensate}}

\subsection{Gross-Pitaevskii description}

The Bose-Einstein condensation of a cloud of $N$ identical bosonic
atoms occurs when their temperature gets below the critical condensation
temperature $T_{c}$. It manifests itself by the macroscopic occupation
of a single mode by the atoms~\citep{Pethick2008,Castin2025}. At
very low temperature compared to the critical temperature, the occupation
of the condensate mode is nearly 100\%.

There is nonetheless a residual fraction of the atoms that are not
in the condensate mode, due to both the non-zero temperature and the
interactions between the atoms. In usual conditions, the non-condensate
fraction originates mostly from the non-zero temperature $T$, and
is given for a harmonic trap by $\left(T/T_{c}\right)^{3}$~\citep{Pethick2008}.
In the experimental situations considered in this study, $T_{c}\sim100\,\text{nK}$
and $T\sim10\text{ nK}$, so that the non-condensate fraction is about
$10^{-3}$. Up to this accuracy, it is safe to make the approximation
that all atoms are described by exactly the same one-body wave function
$\psi(\bm{r},t)$ corresponding to the condensate mode.

We are therefore in the same situation as in Section~\ref{sec:arrivals-of-isolated-particle}
where the predictions of Eq.~(\ref{eq:densityFlux}-\ref{eq:backwardFlux})
are readily applicable once the condensate wave function $\psi(\bm{r},t)$
is known. The only difference is that $\psi(\bm{r},t)$ is not the
solution of the standard Schrödinger equation (\ref{eq:SchroedingerEq})
for an isolated particle, because the interatomic interactions strongly
affect the wave function. Instead, the condensate wave function follows
to a very good accuracy the Gross-Pitaevskii equation~\citep{Pethick2008},
\begin{equation}
i\hbar\dot{\psi}(\bm{r}\!,\!t)\!=\!\left(\!-\frac{\hbar^{2}\nabla^{2}}{2m}+V(\bm{r}\!,\!t)+\frac{4\pi\hbar^{2}a}{m}N\left|\psi(\bm{r}\!,\!t)\right|^{2}\!\right)\psi(\bm{r}\!,\!t),\label{eq:Gross-Pitaevskii-equation}
\end{equation}
where $V(\text{\ensuremath{\bm{r}}},t)$ is the trapping potential,
and $a$ is the scattering length characterising the strength of interactions
between atoms at small densities and temperatures. Stable condensates
usually require repulsive interactions, corresponding to a positive
scattering length $a$. As previously, it will be assumed that the
initial trapping potential $V(\bm{r},0)\eqdef V(\bm{r})$ is of the
harmonic form of Eq.~(\ref{eq:HarmonicTrap}), which is generally
the case for the bottom of commonly used traps, and that it is instantaneously
switched off at $t=0$, such that $V(\bm{r},t)=0$ for $t>0$.

The effect of interactions is particularly important for the initial
state $\psi_{0}$, which is obtained by setting $\psi(\bm{r},t)=\psi_{0}(\bm{r})e^{-i\frac{\mu}{\hbar}t}$
in Eq.~(\ref{eq:Gross-Pitaevskii-equation}), yielding the stationary
equation

\begin{equation}
\mu\psi_{0}(\bm{r})\!=\!\left(\!-\frac{\hbar^{2}\nabla^{2}}{2m}+V(\bm{r})+\frac{4\pi\hbar^{2}a}{m}N\left|\psi_{0}(\bm{r})\right|^{2}\!\right)\psi_{0}(\bm{r})\label{eq:stationary-GP-equation}
\end{equation}
where $\mu$ is the chemical potential fixed by the total number $N$
of atoms, through the normalisation condition,
\begin{equation}
\int\left|\psi_{0}(\bm{r})\right|^{2}d^{3}\bm{r}=1.\label{eq:Normalisation}
\end{equation}

The repulsive effect of the interactions significantly broadens the
spatial extent of the wave function $\psi_{0}$ with respect to the
solution for an isolated atom. It becomes less prominent as the cloud
expands, since the density gradually decreases and the influence of
the non-linear term in Eq.~(\ref{eq:Gross-Pitaevskii-equation})
gets weaker. Nevertheless, it makes the motions along the three directions
$x$, $y$, $z$ interdependent, unlike the non-interacting case discussed
in Section~\ref{sec:arrivals-of-isolated-particle}. It is therefore
necessary to solve the full three-dimensional evolution of the system.
We thus solve Eq.~(\ref{eq:stationary-GP-equation}) and then Eq.~(\ref{eq:Gross-Pitaevskii-equation})
both numerically without any approximation by discretizing space and
time on a grid and evaluating the spatial derivatives by finite differences.

The initial state is obtained by first calculating the ground state
of the stationary equation (\ref{eq:stationary-GP-equation}) without
interactions ($a=0$), and then finding the corresponding interacting
stationary state using the convergent iterative method of Dion-Cancès~\citep{Dion2007}.

The time evolution is then obtained by propagating Eq.~(\ref{eq:Gross-Pitaevskii-equation})
using the rescaling method of Bradley et al.~\citep{Bradley2022}.
It consists in using rescaled coordinates $R_{i}=r_{i}/\lambda_{i}(t)$
such that the scaling factors $\lambda_{i}(t)$ grow with the size
of the cloud. This way, the coordinate system is always suited to
the size of the expanding cloud. The system is thus described by the
set of coupled equations
\begin{align}
\ddot{\lambda_{i}} & =\frac{1}{\lambda_{i}}\frac{1}{ma_{i}^{2}}\int d^{3}\bm{R}\left(\frac{gN}{2}\frac{\left|\phi\right|^{4}}{\lambda_{x}\lambda_{y}\lambda_{z}}+\frac{\hbar^{2}}{m\lambda_{i}}\left|\partial_{i}\phi\right|^{2}\right)\label{eq:TimeEvolution1}\\
i\hbar\dot{\phi} & =\left[\sum_{i}\left(-\frac{\hbar^{2}}{2m}\frac{\partial_{i}^{2}}{\lambda_{i}}+\frac{m}{2}R_{i}^{2}\ddot{\lambda}_{i}\lambda_{i}\right)+\frac{gN\left|\phi\right|^{2}}{\lambda_{x}\lambda_{y}\lambda_{z}}\right]\phi\label{eq:TimeEvolution2}
\end{align}
for the scaling factors $\lambda_{i}(t)$ and the wave function in
rescaled coordinates,
\begin{equation}
\phi(\bm{R},t)\eqdef\psi(\{R_{i}\lambda_{i}(t)\},t)\prod_{j}\lambda_{j}(t)^{1/2}e^{-i\frac{m}{2\hbar}R_{j}^{2}\dot{\lambda}_{j}(t)\lambda_{j}(t)},\label{eq:RescaledWaveFunction}
\end{equation}
with the initial conditions $\lambda_{i}(0)=1$, $\dot{\lambda}_{i}(0)=0$,
and $\phi(\bm{R},0)=\psi_{0}(\bm{r})$. Here, we used the notation
\[
a_{i}\eqdef\left(\int d^{3}\bm{r}r_{i}^{2}\left|\psi_{0}(\bm{r})\right|^{2}\right)^{1/2}
\]
for the spatial extents of the initial state in the $i=x,y,z$ directions.
These equations are solved by performing at each time step a fast
exponentiation of the matrix representing their effective Hamiltonian
multiplied by $\frac{1}{i\hbar}\Delta t$, where $\Delta t$ is the
time step duration.

\subsection{Approximations}

It is useful to compare the numerical results of Eq.~(\ref{eq:stationary-GP-equation})
and Eqs.~(\ref{eq:TimeEvolution1}-\ref{eq:TimeEvolution2}) with
well-known approximations.

\subsubsection{Thomas-Fermi approximation}

The initial density profile can be captured in the limit of large
$Na$ by the Thomas-Fermi approximation, which consists in neglecting
the kinetic term in Eq.~(\ref{eq:stationary-GP-equation}), yielding
\begin{equation}
\left|\psi_{0}(\bm{r})\right|^{2}\approx\left|\psi_{\text{TF}}(\bm{r})\right|^{2}\eqdef\max\left(0,\frac{\mu-V(\bm{r})}{4\pi\hbar^{2}Na/m}\right)\label{eq:TF-approximatiom}
\end{equation}
where the chemical potential is found from Eq.~(\ref{eq:HarmonicTrap})
and Eq.~(\ref{eq:Normalisation}) to be~\citep{Pethick2008},
\begin{equation}
\mu=\frac{1}{2}\hbar\bar{\omega}\left(15Na/\bar{\sigma}\right)^{2/5}\label{eq:chemical-potential}
\end{equation}
where $\bar{\omega}\eqdef\left(\omega_{x}\omega_{y}\omega_{z}\right)^{1/3}$
and $\bar{\sigma}\eqdef\sqrt{\hbar/(m\bar{\omega})}$. In this approximation,
the initial integrated densities 
\begin{equation}
n_{i}(r_{i},0)\eqdef\int dr_{j}dr_{k}\left|\psi_{0}(\bm{r})\right|^{2}\label{eq:initialIntegratedDensity}
\end{equation}
 are given by
\begin{equation}
n_{i}(r_{i},0)=N\frac{15}{16}\frac{1}{R_{i}}\text{max}\left[0,1-\left(\frac{r_{i}}{R_{i}^{\text{TF}}}\right)^{2}\right]^{2}\label{eq:TF-integrated-density}
\end{equation}
where 
\begin{equation}
R_{i}^{\text{TF}}\eqdef\sqrt{\frac{\mu}{\frac{1}{2}m\omega_{i}^{2}}}\label{eq:Thomas-Fermi-radius}
\end{equation}
is the Thomas-Fermi radius in the $i$ direction. 

\subsubsection{Scaling approximation}

The expansion dynamics can be captured to a great extent by assuming
that the rescaled function $\phi$ is time-independent, so that only
the scaling factors $\lambda_{i}(t)$ account for the dynamics:
\begin{equation}
\psi(\bm{r},t)\approx\phi\left(\left\{ \frac{r_{i}}{\lambda_{i}(t)}\right\} \right)\frac{e^{i\frac{m}{2\hbar}\sum_{j}r_{j}^{2}\frac{\dot{\lambda}_{j}(t)}{\lambda_{j}(t)}}}{\prod_{j}\lambda_{j}(t)^{1/2}}.\label{eq:Scaling-approximation}
\end{equation}
 In that case, the integrated densities
\begin{equation}
n_{i}(r_{i},t)\eqdef\int dr_{j}dr_{k}\left|\psi(\bm{r},t)\right|^{2}\label{eq:IntegrateDensity}
\end{equation}
satisfy the scaling equation,
\begin{equation}
n_{i}(r_{i},t)=\frac{1}{\lambda_{i}(t)}n_{i}\left(\frac{r_{i}}{\lambda_{i}(t)},0\right).\label{eq:scalingEquation}
\end{equation}

Note that the non-interacting Gaussian wave packet of Sect.~\ref{sec:arrivals-of-isolated-particle}
exactly satisfies the scaling approximation with the scaling factors
$\lambda_{i}(t)=\sqrt{1+\omega_{i}^{2}t^{2}}$ --- see Eq.~(\ref{eq:Scaling-Factor}).

\subsubsection{Scaling Thomas-Fermi approximation\label{subsec:Scaling-Thomas-Fermi-approximation}}

The scaling Thomas-Fermi approximation corresponds to the conjonction
of the two preceding approximations, i.e. the initial state is assumed
to be in the Thomas-Fermi limit Eq.~(\ref{eq:TF-approximatiom})
and the dynamics is assumed to follow the scaling approximation. In
that case, the scaling factors $\lambda_{i}$ are found to obey the
equations~\citep{Castin1996},
\begin{equation}
\ddot{\lambda}_{i}(t)=\frac{\omega_{i}^{2}}{\lambda_{i}(t)\lambda_{x}(t)\lambda_{y}(t)\lambda_{z}(t)},\label{eq:Thomas-Fermi-equations-for-lambdas}
\end{equation}
which, for a cylindrical symmetry, reduce to
\begin{align}
\ddot{\lambda_{\perp}}(t) & =\frac{\omega_{\perp}^{2}}{\lambda_{\perp}(t)^{3}\lambda_{\parallel}(t)},\label{eq:Eq-lambda-xy}\\
\ddot{\lambda}_{\parallel}(t) & =\frac{\omega_{\parallel}^{2}}{\lambda_{\perp}(t)^{2}\lambda_{\parallel}(t)^{2}},\label{eq:Eq-lambda-z}
\end{align}
where $\parallel$ designates the direction of the axis of symmetry,
and $\perp$ designates any transverse direction.

Although the above equations Eqs.~(\ref{eq:Eq-lambda-xy}-\ref{eq:Eq-lambda-z})
will be solved numerically in Sec.~\ref{sec:Results}, one can consider
different analytical limits: the ``pancake'' limit ($\omega_{\parallel}\gg\omega_{\perp}$),
the symmetric case ($\omega_{\parallel}=\omega_{\perp}$), and the
``cigar'' limit ($\omega_{\parallel}\ll\omega_{\perp}$). For large
times when $\lambda_{i}(t)\gg1$, one finds that the scaling factors
$\lambda_{i}(t)$ end up growing linearly (or almost linearly) with
time. Specifically, one obtains the following asymptotic behaviours~\citep{Castin1996},
\begin{align}
\dot{\lambda}_{\parallel}\! & \sim\!\sqrt{2}\omega_{\parallel}\,;\,\dot{\lambda}_{\perp}\!\!\sim\frac{\omega_{\perp}^{2}\ln\omega_{\parallel}t}{\sqrt{2}\omega_{\parallel}}\quad\text{for }\omega_{\perp}\!\ll\omega_{\parallel}\text{ (pancake)}\label{eq:Pancake-limit}\\
\dot{\lambda}_{\parallel} & \sim s_{1}\omega_{\parallel}\quad;\quad\dot{\lambda}_{\perp}\sim s_{1}\omega_{\perp}\quad\text{ for }\omega_{\perp}=\omega_{\parallel}\text{ (sym.)}\label{eq:Symmetric-limit}\\
\dot{\lambda}_{\parallel} & \sim\frac{\pi}{2}\frac{\omega_{\parallel}^{2}}{\omega_{\perp}}\quad;\quad\dot{\lambda}_{\perp}\sim\omega_{\perp}\quad\text{ for }\omega_{\perp}\gg\omega_{\parallel}\text{ (cigar)}\label{eq:Cigar-limit}
\end{align}
where $s_{1}\approx0.81649$. One can see that in most cases, the
rate of expansion $\dot{\lambda}_{i}$ in direction $i$ is proportional
to the corresponding frequency $\omega_{i}$ in that direction, just
as in the non-interacting case. However, in the very loose trapping
directions, the rate $\dot{\lambda}_{i}$ depends on both frequencies
$\omega_{\parallel}$ and $\omega_{\perp}$, and can be adjusted to
much smaller values than the typical experimental values of these
frequencies by varying their ratio. This is one of the strongest qualitative
differences between the expansion of isolated atoms and that of interacting
condensed atoms. In Sec.~\ref{sec:Results}, we will focus on a cigar-shaped
condensate.

\subsubsection{Arrival-time distribution}

In the scaling Thomas-Fermi approximation, one can analytically calculate
the first three arrival-time distributions Eq.~(\ref{eq:densityFlux}-\ref{eq:backwardFlux})
for arrivals along direction $i$ on a planar surface $S$ perpendicular
to it:
\begin{align}
\Pi(t) & =\alpha\frac{1}{\lambda_{i}(t)}n_{i}\left(\frac{r_{i}}{\lambda_{i}(t)},0\right)\label{eq:Scaling-DensityFlux}\\
F(t) & =\frac{\hbar}{mR_{i}^{\text{TF}}}\frac{r_{i}}{R_{i}^{\text{TF}}}\frac{1}{\lambda_{i}(t)^{2}}B(t)n_{i}(r_{i},t)\label{eq:Scaling-ProbabilityFlux}\\
F_{-}(t) & =\frac{\hbar}{mR_{i}^{\text{TF}}}\frac{r_{i}}{R_{i}^{\text{TF}}}\frac{1}{\lambda_{i}(t)^{2}}\left(B(t)+A(r_{i},t)\right)n_{i}(r_{i},t)\label{eq:scaling-BackwardFlux}
\end{align}
with
\begin{align}
B(t) & =\frac{m\left(R_{i}^{\text{TF}}\right)^{2}}{\hbar}\dot{\lambda}_{i}(t)\lambda_{i}(t)\label{eq:B}\\
A(r_{i},t) & =2\left[1-\left(\frac{r_{i}}{\lambda_{i}(t)R_{i}^{\text{TF}}}\right)^{2}\right]^{-1}\label{eq:A}
\end{align}
Note that Eqs.~(\ref{eq:Scaling-DensityFlux}-\ref{eq:scaling-BackwardFlux})
are also exactly satisfied in the non-interacting Gaussian case of
Section~\ref{sec:arrivals-of-isolated-particle} with $R_{i}^{\text{TF}}=\sigma_{i}$,
$\lambda_{i}(t)=\sqrt{1+\omega_{i}^{2}t^{2}}$ and $A(x_{i})=1$,
see Eqs.~(\ref{eq:Gaussian-DensityFlux}-\ref{eq:Gaussian-BackwardFlux}).
Unfortunately, the scaling Thomas-Fermi approximation does not appear
to yield any analytical expression for the Kijowski flux $\Pi_{\text{K}}$.

\subsection{Results for a metastable helium condensate\label{sec:Results}}

\begin{figure*}
\includegraphics[viewport=30bp 0bp 450bp 286bp,clip,height=6.3cm]{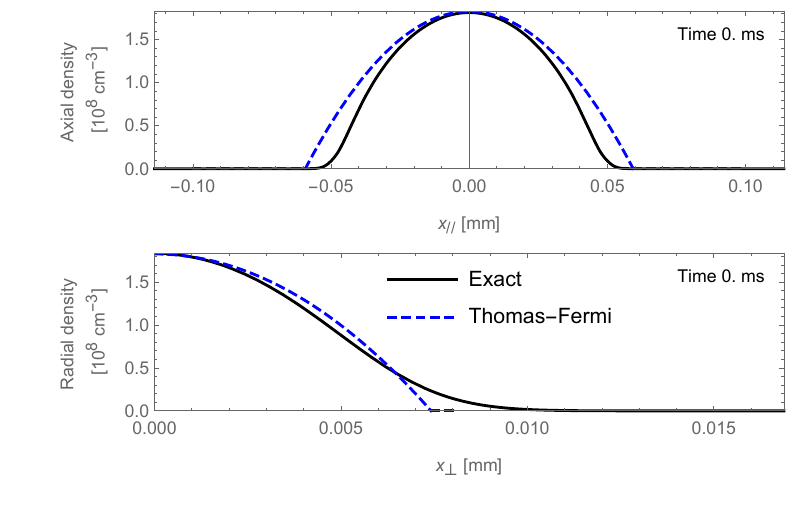}\includegraphics[viewport=40bp 0bp 450bp 286bp,clip,height=6.3cm]{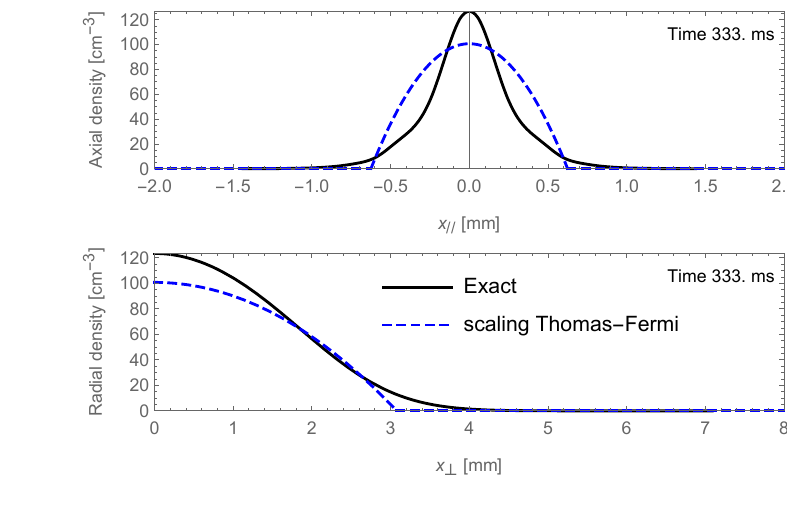}

\caption{\label{fig:Density-cuts}Density cuts along the axial ($x_{\parallel}$)
and transverse ($x_{\perp}$) directions through a cigar-shaped condensate
of $10^{4}$ helium atoms initially held in a trap with axial frequency
$\omega_{\parallel}=2\pi\times25\text{ Hz}$ and transverse frequency
$\omega_{\perp}=2\pi\times200\text{ Hz}$. Left panel: initial state.
Right panel: after expanding for 0.333 s. Solid curves: numerical
solutions of the exact theory, Eq.~(\ref{eq:stationary-GP-equation})
(left) and Eqs.~(\ref{eq:TimeEvolution1}-\ref{eq:TimeEvolution2})
(right). Dashed curves: scaling Thomas-Fermi approximation Eqs.~(\ref{eq:TF-approximatiom}),
(\ref{eq:Scaling-approximation}) and (\ref{eq:Eq-lambda-xy}-\ref{eq:Eq-lambda-z}).}
\end{figure*}
\begin{figure}
\includegraphics[width=8cm]{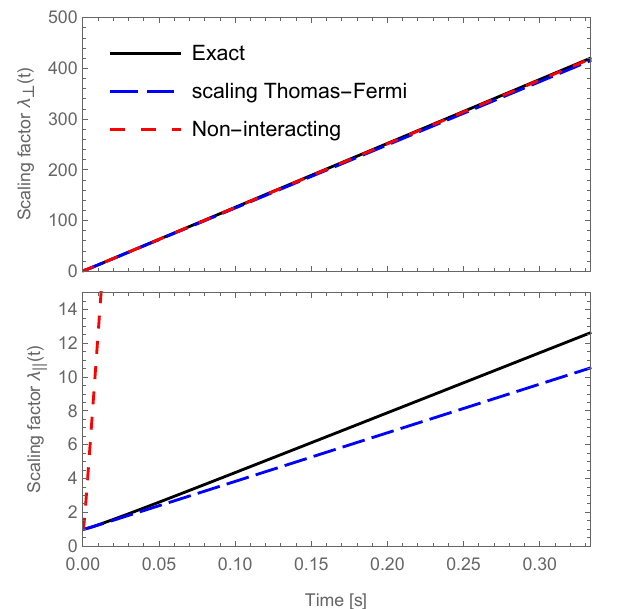}

\caption{\label{fig:Scaling-factors}Scaling factors $\lambda_{\perp}(t)$
and $\lambda_{\parallel}(t)$ of the condensate in the transverse
and axial directions. Solid lines: numerical solutions of the exact
theory, Eqs.~(\ref{eq:TimeEvolution1}-\ref{eq:TimeEvolution2}).
Dashed lines: numerical solutions of the scaling Thomas-Fermi approximation,
Eqs.~(\ref{eq:Eq-lambda-xy}-\ref{eq:Eq-lambda-z}). Dotted lines:
non-interacting case, Eq.~(\ref{eq:Scaling-Factor}).}

\end{figure}
\begin{figure}
\includegraphics[width=8cm]{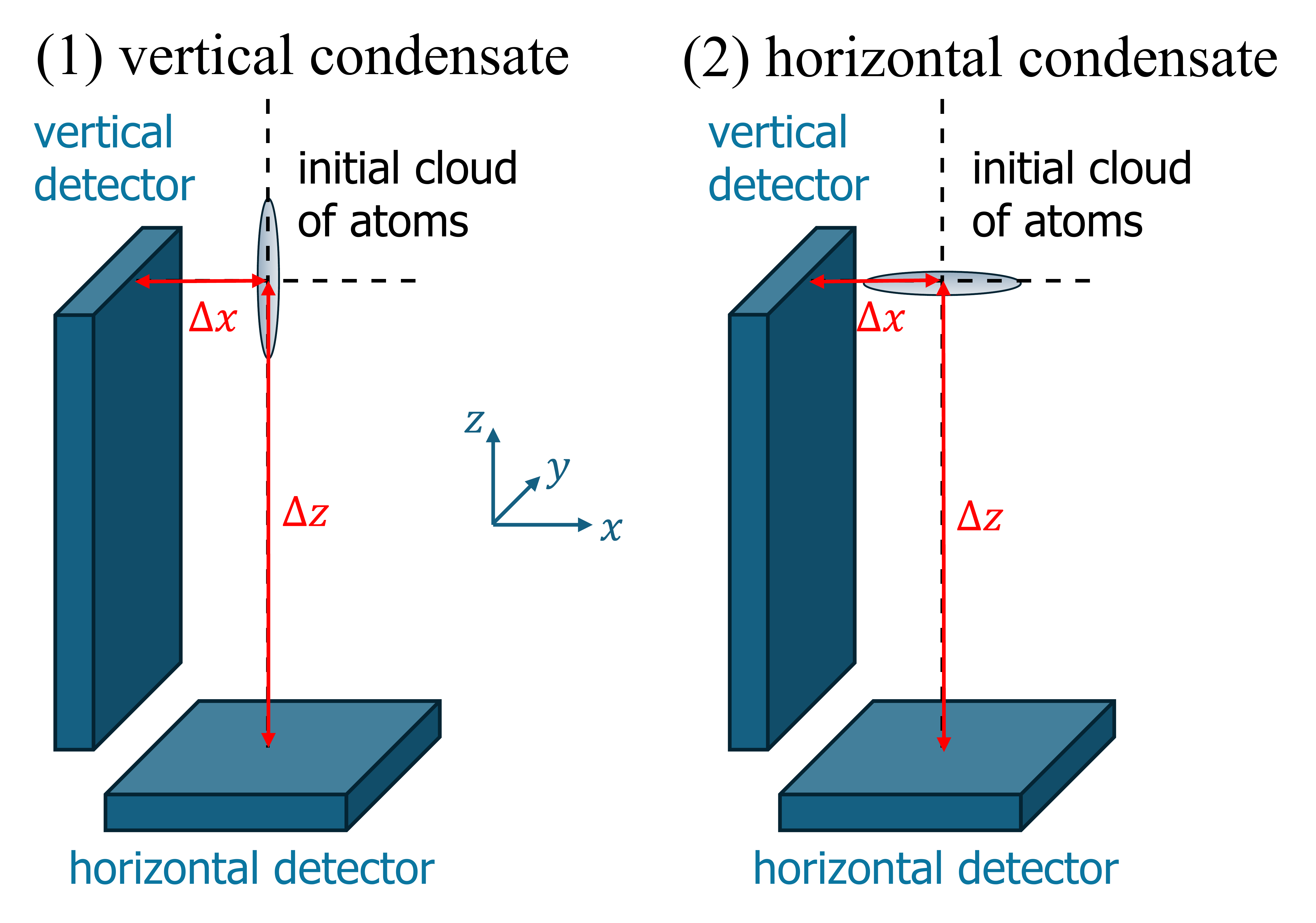}

\caption{\label{fig:setup}Schematic setup for detecting the time of flight
of a Bose-Einstein condensate on a horizontal detector and a vertical
detector. The condensate is cigar-shaped and oriented vertically (1)
or horizontally (2).}
\end{figure}

\subsubsection{Setup}

To describe a typical experiment, we consider a condensate of $10^{4}$
metastable helium-4 atoms ($a=7.5\,\text{nm}$) held in a harmonic
trap of cylindrical symmetry, with axial trapping frequency $\omega_{\parallel}=2\pi\times25\,\text{Hz}$
and transverse frequency $\omega_{\perp}=2\pi\times200\text{ Hz}$.
This results in a cigar-shaped Bose-Einstein condensate as an initial
state. The density cuts along the symmetry axis and transverse direction
are obained from the numerical solution of Eq.~(\ref{eq:stationary-GP-equation})
and shown in the left panel of Fig.~\ref{fig:Density-cuts}. One
can see that the Thomas-Fermi approximation (dashed curves) qualitatively
captures the exact density profile (solid curves).

At time $t=0$, the condensate is released from the trap and starts
expanding. Its wave function is computed by numerically solving Eqs.~(\ref{eq:TimeEvolution1}-\ref{eq:TimeEvolution2}).
The right panel of Fig.~\ref{fig:Density-cuts} shows the density
cuts through the condensate after it has expanded for 0.333~s. One
can see that the scaling approximation captures the rapid expansion
of the condensate, especially in the transverse direction in which
it expands by two orders of magnitude. Nevertheless, the agreement
with the exact density profiles remains qualitative, especially in
the axial direction where there is a significant deformation of the
original profile.

Figure~\ref{fig:Scaling-factors} shows the scaling factors $\lambda_{\perp}(t)$
and $\lambda_{\parallel}(t)$ of the condensate in the transverse
and axial directions, as a function of time. One can see that the
condensate grows essentially linearly with time, as discussed in Sec.~\ref{subsec:Scaling-Thomas-Fermi-approximation}.
This expansion is relatively well captured by the scaling Thomas-Fermi
approximation, especially in the transverse direction, for which the
expansion also remains close to that of a non-interacting system at
frequency $\omega_{\perp}$. In contrast, in the axial direction,
the expansion rate is much slower than the expansion rate $\omega_{\parallel}$
of the non-interacting case, in agreement with the discussion in Sec.~\ref{subsec:Scaling-Thomas-Fermi-approximation}.
It would correspond to the effective trapping frequency $\omega_{\text{eff}}\approx2\pi\times5.65\,\text{Hz}$
of a non-interacting gas.

As previously, the arrivals of the cloud can be detected on a horizontal
detector (parallel to the $xy$ plane) and a vertical detector (parallel
to the $yz$ plane). As we saw in Sec.~\ref{subsec:Vertical-arrivals},
the horizontal detector is not useful for distinguishing different
predictions of the arrival-time distribution because they are all
dominated by gravity. In the following, we shall therefore focus on
arrivals on the vertical detector, which are not affected by gravity.

We will consider two different orientations of the trap: an orientation
(1) for which the axis of the trap is vertical (parallel to the $z$-axis),
and another (2) for which the axis of the trap is horizontal and parallel
to the $x$-axis, as shown in Fig.~\ref{fig:setup}.

\begin{figure}

\includegraphics[width=8cm]{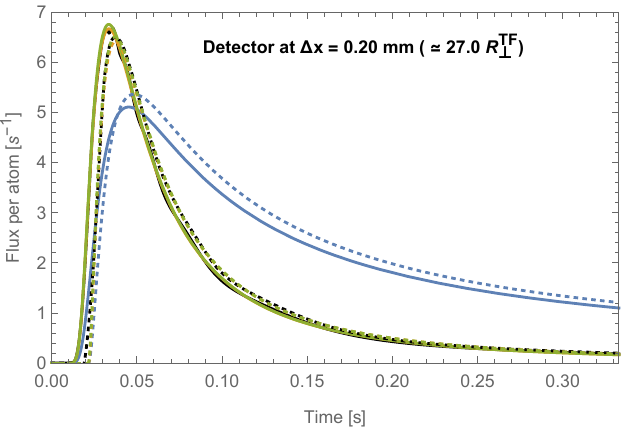}

\includegraphics[width=8cm]{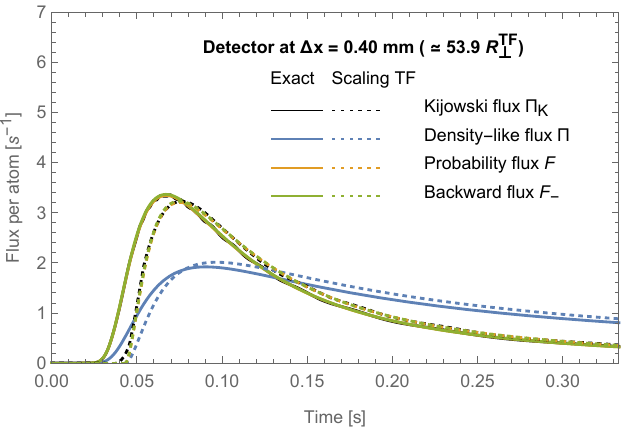}

\caption{\label{fig:CondensateOrientationA}Distribution of times of arrivals
of a metastable helium condensate in orientation (1) onto a vertical
detector at two different distances from the centre of the trap, $\Delta x=0.20,\,0.40\,\text{mm}$
(top, bottom). The solid curves represent the numerical results for
different predictions, Eqs.~(\ref{eq:densityFlux}-\ref{eq:KijowskiDistribution}),
and the dashed curves are the corresponding results in the scaling
Thomas-Fermi (TF) approximation. The density-like flux is shown here
with the coefficient $\alpha$ set to the arbitrary value of 3~mm/s.
Mind that the quantity plotted here is the \emph{flux per atom }for
easier comparison with the isolated-atom case ; this flux should be
multiplied by the total number of atoms $N=10^{4}$ to obtain the
total flux.}
\end{figure}

\begin{figure}
\includegraphics[width=8cm]{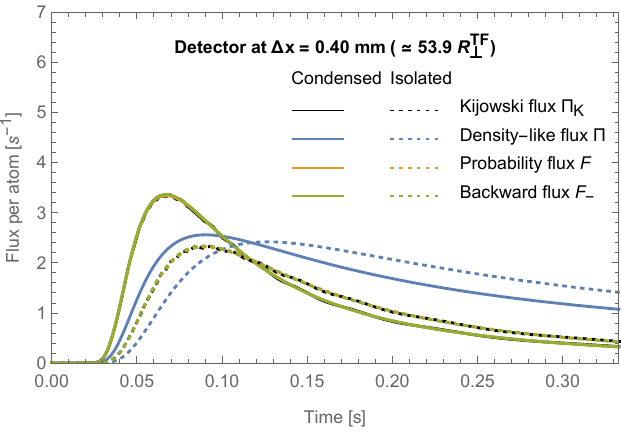}

\caption{\label{fig:Comparison-transverse}Distribution of times of arrivals
of a metastable helium condensate in orientation (1) onto a vertical
detector at the distance $\Delta x=0.40\,\text{mm}$ from the centre
of the trap (same as bottom panel of Fig.~\ref{fig:CondensateOrientationA}).
The solid curves show the numerical results of Eqs.~(\ref{eq:densityFlux}-\ref{eq:KijowskiDistribution})
and the dashed curves show the isolated-atom results Eqs.~(\ref{eq:Gaussian-DensityFlux}-\ref{eq:Gaussian-KijowskiFlux})
for comparison.}
\end{figure}

\subsubsection{Vertical trap axis: transverse arrivals\label{subsec:Vertical-trap-axis:}}

We first consider the orientation (1) of the condensate, for which
the trap axis is vertical. In this configuration, the detector monitors
the arrivals along a transverse direction of the condensate. 

The different predictions for the arrival-time distribution are shown
in Fig.~\ref{fig:CondensateOrientationA} for two distances of the
detector from the trap. We take 0.2 mm as the smallest distance, which
is about 25 times larger than the initial radius of the condensate
in the transverse direction, so that the detector is not expected
to affect the initial state.

One can see that the arrival-time distributions are relatively well
captured by the scaling Thomas-Fermi approximation, shown by the dashed
curves in Fig.~\ref{fig:CondensateOrientationA}. On the other hand,
Fig.~\ref{fig:Comparison-transverse} shows that they significantly
differ from the isolated-atom arrival-time distributions. Since the
expansion dynamics is nearly the same for both condensate atoms and
isolated atoms --- see top of Fig.~\ref{fig:Scaling-factors} ---,
these quantitative differences should be ascribed mostly to the difference
in the initial density profiles. In spite of these quantitative differences,
the distributions remain qualitatively very similar. In particular,
all predictions are nearly indistinguishable, except the density-like
flux, as previously found for isolated atoms. 

Stronger differences between predictions could be obtained by placing
the detector closer to the trap, but this could be technically challenging.
Another option would be to use lower trapping frequencies, but there
are also technical limitations on achieving and trapping a condensate
in a loose trap. 

\begin{figure}

\includegraphics[width=8cm]{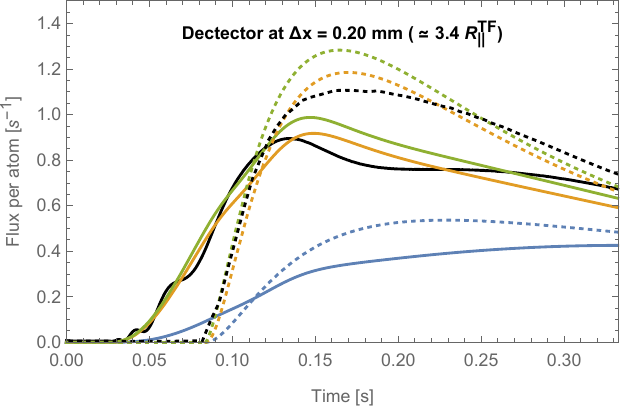}

\includegraphics[width=8cm]{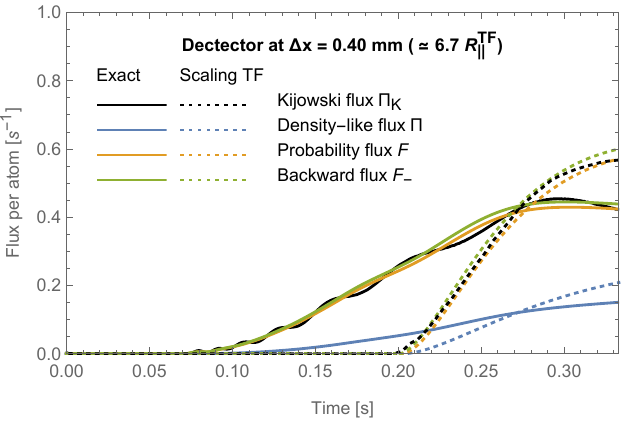}

\caption{\label{fig:CondensateOrientationB}Distribution of times of arrivals
of a metastable helium condensate in orientation (2) onto a vertical
detector at two different distances from the centre of the trap, $\Delta x=0.20,\,0.40\,\text{mm}$
(top, bottom). Conventions are the same as in Fig.~\ref{fig:CondensateOrientationA}
and $\alpha=0.4\,$mm/s.}
\end{figure}

\begin{figure}
\includegraphics[width=8cm]{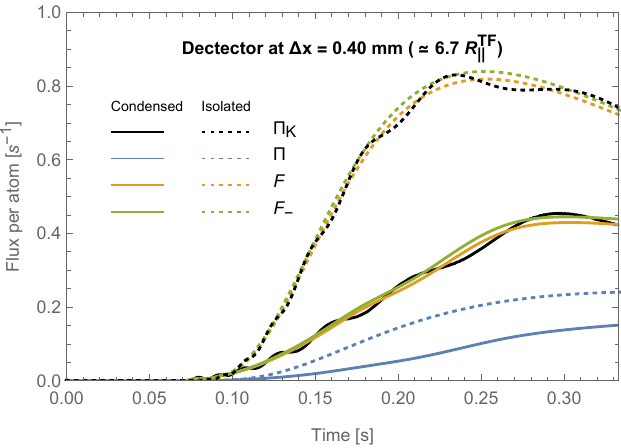}

\caption{\label{fig:Comparison-axial}Distribution of times of arrivals of
a metastable helium condensate in orientation (2) onto a vertical
detector at the distance $\Delta x=0.40\,\text{mm}$ from the centre
of the trap (same as bottom panel of Fig.~\ref{fig:CondensateOrientationB}).
The solid curves show the numerical results of Eqs.~(\ref{eq:densityFlux}-\ref{eq:KijowskiDistribution})
and the dashed curves show the isolated-atom results Eqs.~(\ref{eq:Gaussian-DensityFlux}-\ref{eq:Gaussian-KijowskiFlux})
for comparison.}

\end{figure}

\subsubsection{Horizontal trap axis: axial arrivals}

We now consider the orientation (2) of the initial cloud, for which
the trap axis is horizontal and perpendicular to the vertical detector.
In that configuration, the vertical detector monitors arrivals in
the axial direction of the trap. As we saw in Sec.~\ref{subsec:Scaling-Thomas-Fermi-approximation}
and in Fig.~\ref{fig:Scaling-factors}, due to the elongation of
the trap, the expansion in the axial direction is now much slower
than the expansion of non-interacting atoms.

The corresponding arrival-time distributions are shown in Fig.~\ref{fig:CondensateOrientationB},
for the same detector distances as previously. It can be seen that
the differences between predictions are now more pronounced. This
underscores a key advantage of interacting condensates over non-interacting
atoms: by controlling the expansion of the cloud at a much lower rate
than the trapping frequencies thanks to a squeezed initial geometry,
one can solve the issue mentioned at the end of Sec\@.~\ref{subsec:Vertical-trap-axis:}.
Indeed, this enables to probe the regime where predictions differ
more strongly, while the physical frequencies remain large enough
to trap the atoms..

One can see that the scaling Thomas-Fermi approximation (dashed curves)
is rather poor here. This is related to the fact that, unlike the
transverse profile, the axial profile is significantly deformed ---
see Fig.~(\ref{fig:Density-cuts}) --- and the scaling approximation
consequently yields an inaccurate rate of expansion --- see Fig\@.~(\ref{fig:Scaling-factors}).

Figure~\ref{fig:Comparison-axial} shows that the numerical results
are also different from the isolated-atom case, since now, not only
the initial density profile is different, but the expansion dynamics
is also slower. Nevertheless, the condensate and isolated-atom cases
are again qualitatively similar.

As in the isolated-atom case, significant differences between predictions
appear mostly for small detector distances. The smallest considered
distance, $\Delta x=0.20\,\text{mm}$, is only about 4 times the axial
radius of the initial cloud, which means that the detector is sitting
relatively close to the initial cloud. Although the discrepancies
between predictions are reduced for larger distances of the detector,
the Kijowski flux now clearly exhibits the oscillatory pattern found
previously in the isolated-atom case, which is still noticeable at
the distance $\Delta x=0.40\,\text{mm}$ (about 8 times the axial
initial radius). This confirms that such oscillations can also be
present in an interacting condensate, opening the possibility to discriminate
the Kijowski flux from other predictions~\citep{Naidon2025a}.

\section{Conclusion\label{sec:Conclusion}}

In this work, the arrival times of an atomic Bose-Einstein condensate
has been investigated in the short-distance regime. Compared to isolated
particles, Bose-Einstein condensates allow to probe at once the arrivals
of many particles in the same conditions, enhancing the statistics
by several orders of magnitude. Although the interactions between
atoms bring strong quantitative differences with respect to the arrival
times of isolated particles, we found no major qualitative difference.
In particular, discrepancies found between various predictions of
arrival-time distribution for an isolated particle remain in the case
of a Bose-Einstein condensate.

Moreover, we identified favourable conditions for the experimental
discrimination of these predictions. Specifically, we find it is advantageous
to measure the arrivals from a loose direction of an asymmetric trap.
In such conditions, it appears that both the density-like distribution
predicted by the quantum clock approach~\citep{Giovannetti2015,Maccone2020,Roncallo2023}
and the Kijowski distribution predicted by the axiomatic approach~\citep{Kijowski1974}
and other works~\citep{Aharonov1961,Allcock1969b,Grot1996,Delgado1997,Anastopoulos2006}
can be discriminated from other predictions. In particular, the Kijowski
distribution was shown to exhibit distinctive oscillations absent
from other predictions.

These results can serve as a basis for future experiments with cold
atoms. In a companion article~\citep{Naidon2025a}, we build on these
results to make a concrete experimental proposal.\medskip{}

\begin{acknowledgments}P. N. acknowledges support from the JSPS Grants-in-Aid
for Scientific Research on Innovative Areas (No.JP23K03292). L. H.
is supported by the RIKEN special postdoctoral researcher program.
D. B. acknowledges founding from QuantERA Grant No. ANR-22-QUA2-000801
(MENTA), ANR Grant No. 20-CE-47-0001-01 (COSQUA), and the LabEx PALM
(No. ANR-10-LABX-0039PALM). \end{acknowledgments}

\bibliographystyle{apsrev4-2}
\bibliography{paper46}

\end{document}